\documentclass[aps,pra,twocolumn,noshowpacs,noshowkeys]{revtex4-1}

\usepackage[colorlinks=true,
  citecolor  =blue,
  anchorcolor=blue,
  urlcolor   =blue,linkcolor=purple
  ]{hyperref}

\usepackage{newtxtext,newtxmath}

\usepackage[mathscr]{eucal}

\DeclareMathOperator{\sgn} {sgn}

\DeclareMathOperator{\nint} {nint}

\usepackage{graphicx,pstricks}

\begin{document}

\title{Phase-amplitude formalism for ultra-narrow shape resonances}

\date{\today}

\author{I. Simbotin}
\author{D. Shu}
\author{R. C\^ot\'e}

\affiliation{Department of Physics, University of Connecticut,
   2152 Hillside Rd., Storrs, CT 06269-3046, USA}

\begin{abstract}

We apply Milne's phase-amplitude representation [W. E. Milne,
  Phys.~Rev.~\textbf{35}, 863 (1930)] to a scattering problem
involving disjoint classically allowed regions separated by a barrier.
Specifically, we develop a formalism employing different sets of
amplitude and phase functions\,---\,each set of solutions optimized
for a separate region\,---\,and we use these locally adapted solutions
to obtain the true value of the scattering phase shift and accurate
tunneling rates for ultra-narrow shape resonances.  We show results
for an illustrative example of an attractive potential with a large
centrifugal barrier.

\end{abstract}



\maketitle

\section{Introduction}

An integral representation for scattering phase shifts based on the
phase-amplitude formalism was recently derived by the present
authors~\cite{delta_int}.  Although the main result of
Ref.~\cite{delta_int} is fully general, the computational approach was
restricted to a single (infinite) classically allowed region; thus, in
the presence of a barrier, our previous method can only be employed
for energies above the barrier.  We now extend the phase-amplitude
formalism~\cite{milne} to scattering energies below the top of the
barrier, and we provide a  method for characterizing shape
resonances.  We pay special attention to the case of a large barrier
delimiting a deep inner well capable of holding long-lived resonances.
A variety of methods \cite{res-stablz-1970, res-stablz-1993, res-jim,
  sidky-barrier,prl-98-gdb,mrugala-2008} have been developed for
tunneling resonances; however, the regime of ultra-narrow resonances
($\Gamma\lll E_\text{res}$) still presents computational
difficulties~\cite{sidky-barrier, prl-98-gdb, mrugala-2008}.  The
phase-amplitude approach presented in this work overcomes this
obstacle, as it yields the scattering phase shift expressed in terms
of quantities obeying a simple energy dependence and allows the
extraction of highly accurate resonance widths.

Milne's phase-amplitude method~\cite{milne} has a long history and has
been used extensively in atomic physics
\cite{Chris_Greene_QDT_milne_formalism,Chris_Greene_QDT_surface,
  robicheaux,raoult-1988-prl,raoult-1994,phase_amp_John_PRA49,
  Greene_John_PRL81,John_and_Paul,second_order_WKB_wf_correction,plasma-rad,
  crubellier,raoult-2007,fabrikant-2012,olivier-raoult-2013,greene-2018-H2+}.
However, its wealth of advantages is still being explored
\cite{jap-rho-linear,delta_int, cosmo-2018-epjc,
  cosmo-inflat-2018-arxiv}.  In this study we exploit the relationship
between the solutions of the radial Schr\"odinger equation and those
of the envelope equation (which is equivalent with Milne's amplitude
equation).  In particular, we develop an approach for extending the phase
function outside its domain of smoothness, which makes it possible to
combine solutions that are locally adapted in each classically allowed
region and thus bridge them across the barrier.  Making use of our new
results, we can now extend the applicability of the integral
representation in Ref.~\cite{delta_int} to scattering energies below
the top of the barrier, which allows us to analyze ultra-narrow shape
resonances.

This article is organized as follows.  Section~\ref{sec:theory-1}
gives the theoretical description of our phase-amplitude approach,
which makes it possible to separate the background and resonant
contributions to the scattering phase shift; see
Sec.~\ref{sec:bg+res}\@.  The resonance widths are obtained in
Sec.~\ref{sec:Gamma}, and results for an illustrative example are
presented in Sec.~\ref{sec:results}\@.  Concluding remarks are given
in Sec.~\ref{sec:end}.

\section{Theory:  envelope equation approach}
\label{sec:theory-1}

We consider the scattering of two structureless, spinless particles
with a spherically symmetric potential $V(R)$.  The radial
Schr\"odinger equation reads
\begin{equation}\label{eq:radial}
  \Psi'' = U\Psi, \qquad U=2\mu\big(V_{\rm eff}-E\big),
\end{equation}
where $V_{\rm eff}(R)=V(R)+\frac{\ell(\ell+1)}{2\mu R^2}$ is the
effective potential, $\mu$ is the reduced mass of the two particles
undergoing scattering, and $E=\frac{k^2}{2\mu}>0$ is the energy in the
center-of-mass frame.  Atomic units are used throughout.

 \subsection{ The envelope equation}

         \label{sec:rho}

As in our previous work~\cite{delta_int} (see also
Ref.~\cite{schief-linear-rho,leach-2000,jap-rho-linear}), the
Schr\"odinger equation is replaced by the envelope equation,
\begin{equation}\label{eq:rho-linear}
\rho''' = 4U\!\rho' + 2U'\!\rho.
\end{equation}
A particular solution $\rho(R)$ and its corresponding phase
$\theta(R)$ can be  used to parametrize the physical wave function,
\begin{equation}\label{eq:psi=rho-theta}
\psi(R) = \sqrt{\rho(R)} \sin[\theta(R)-\theta(0)],
\end{equation}
and to obtain the scattering phase shift,
\begin{equation}\label{eq:delta=-theta}
\delta_\ell = \ell\textstyle\frac\pi2 -\theta(0).
\end{equation}
This result relies on the smoothness of $\rho(R)$ and $\theta(R)$ in
the asymptotic region, which is ensured using the computational
approach of Ref.~\cite{delta_int}.  Namely, $\rho(R)$ is initialized at
$R=\infty$ according to the asymptotic boundary condition
\[
\rho(R)  \xrightarrow{R\to\infty}  1,
\]
and is propagated inward.  The envelope function $\rho(R)$ is then
used to obtain $\theta(R)$ by integrating
\begin{equation}\label{eq:theta-prime-rho}
\theta'=\frac k \rho.
\end{equation}
The phase function will thus obey the asymptotic behavior
\[
\theta(R)  \xrightarrow{R\to\infty}  kR.
\]

Our main goal is computing the phase $\theta(R)$ at $R=0$, which
yields the phase shift $\delta_\ell$ in Eq.~(\ref{eq:delta=-theta}).
In our previous work~\cite{delta_int} we presented a method suitable
for the case of a single classically allowed region extending to
infinity.  However, if the effective potential $V_\text{eff}(R)$ has a
barrier, and if the scattering energy is below the top of the barrier,
the direct propagation (numerical integration) of the outer phase
$\theta$ into the inner potential well is no longer feasible, as we
explain below.

\begin{figure}[b]
\includegraphics[width=\linewidth]{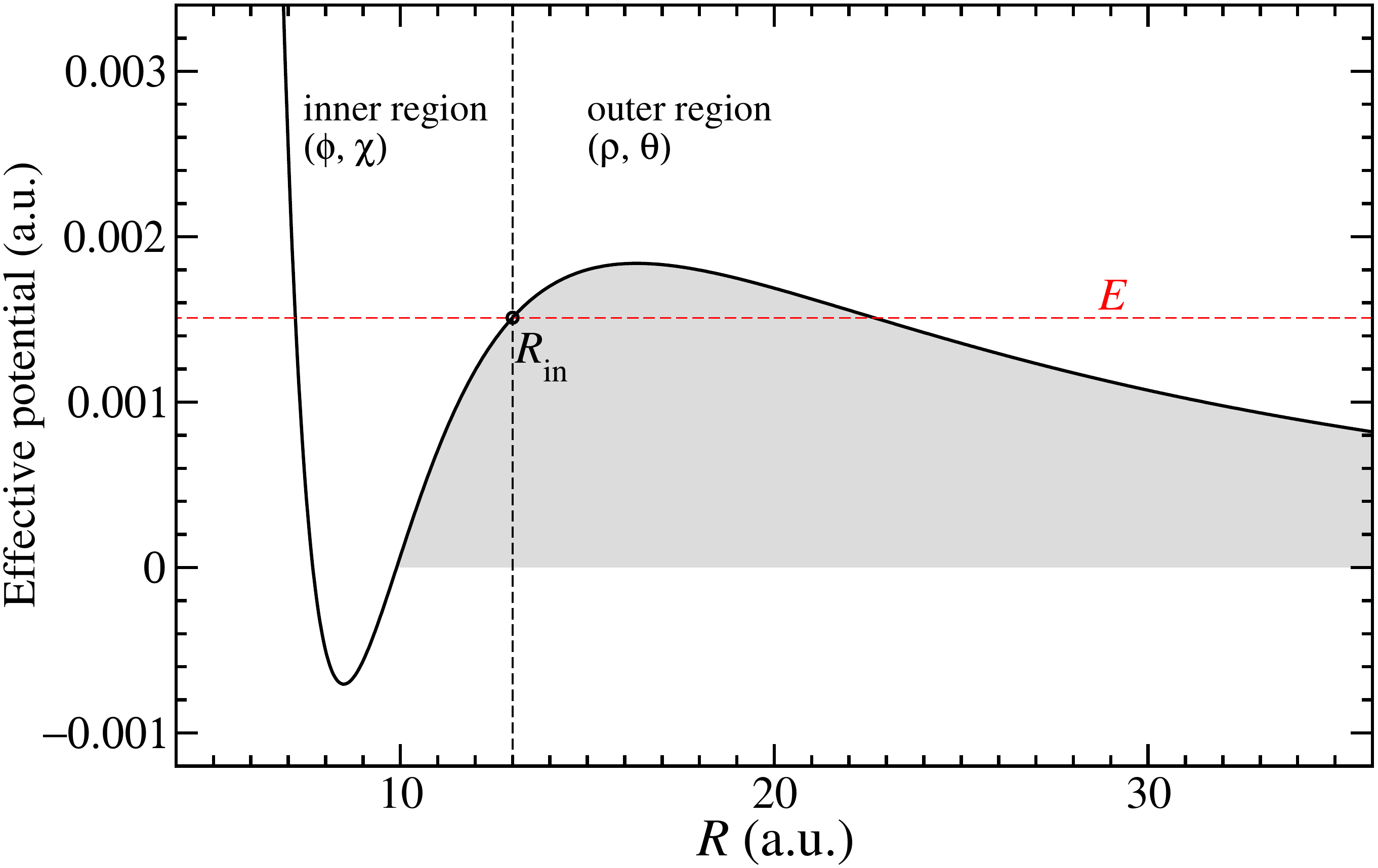}
\caption{
  \label{fig:Veff}
  A representative effective potential which has a sufficiently deep
  well at short range, delimited by a large barrier (indicated by the
  shaded area).  The dashed vertical line at the turning point
  $R_\text{in}$ separates the inner and outer regions.  }
\end{figure}

An example of a potential with a large barrier is depicted in
Fig.~\ref{fig:Veff}.  For energies $0<E<E_\text{top}$, where
$E_\text{top}$ is the height of the barrier, two classically allowed
regions exist, which are separated by the barrier.  We thus divide the
radial domain in two regions, as shown in Fig.~\ref{fig:Veff}.  The
turning point on the inner side of the barrier, $R_{\rm in}(E)$, is
the boundary between the inner and the outer regions.  The latter
includes the classically forbidden region under the barrier and the
entire asymptotic domain.

The outer envelope and phase, $\rho$ and $\theta$, are propagated
inward through the asymptotic region and through the barrier, using
the method we presented in Ref.~\cite{delta_int}.  We remark that the
classically forbidden region under the barrier does not pose any
difficulties.  However, the envelope $\rho(R)$ increases
quasi-exponentially, as $R$ decreases through the barrier region;
thus, $\rho(R_\text{in})$ will be very large.  This can be easily
understood if we write $\rho=f^2+g^2$, where $f$ and $g$ are solutions
of the radial equation which obey the asymptotic behavior
$f(R)\sim\sin(kR)$ and $g(R)\sim\cos(kR)$.  According to their
definition, $f$ and $g$ are linearly independent; hence, one solution
(say, $g$), or both of them, must increase through the barrier, as $R$
decreases towards $R_\text{in}$.  Thus, the dominant solution ($g$)
will dictate the behavior of the envelope \emph{inside the inner
  well}, where we have $\rho(R)=g^2(R)$ to a very good approximation;
consequently, for $R<R_\text{in}$, the envelope has an oscillatory
behavior with (nearly) vanishing minima at the nodes of $g$, and
exceedingly large values at the anti-nodes.  This would cause severe
difficulties if $\theta(R)$ were propagated inside the inner well
($R<R_\text{in}$).  Indeed, when integrating
Eq.~(\ref{eq:theta-prime-rho}), the minima of $\rho$ yield a series of
sharp spikes for the integrand $\frac{k}{\rho(R)}$, which cannot be
handled numerically.  Therefore, the inner region has to be tackled
separately (independently of the outer region), and the two regions
need to be bridged together, in order to obtain $\theta(0)$.

\subsection{  Linear decomposition  of envelope solutions}

        \label{sec:decomp}

As is well known, the general solution of Milne's amplitude
equation can be expressed \cite{pinney, 1976-abc, 1980-abc, 1981-abc,
  Chris_Greene_QDT_surface} in terms of solutions of the radial
equation~(\ref{eq:radial}).  Equivalently, the general solution
of the envelope equation~(\ref{eq:rho-linear}) can be written as
\begin{equation}
  \label{eq:rho=abc-lin-combi}
\rho = a\phi^2 + b\chi^2 +2c\phi\chi,
\end{equation}
where $\phi$ and $\chi$ are linearly independent solutions of
Eq.~(\ref{eq:radial}).  The coefficients
$a$, $b$ and $c$ are free in
general, but they can be chosen to obey the constraint
\begin{equation}\label{eq:abc-constr-k}
\left(ab - c^2\right)W^2 =  k^2,
\end{equation}
with $W$ the Wronskian of $\phi$ and $\chi$.  The constraint above is
directly related to an invariant of the envelope equation, as
explained in Appendix~\ref{app:proof}\@.  We emphasize that $\phi^2$,
$\chi^2$ and $\phi\chi$ are particular solutions of the envelope
equation ; see Ref.~\cite{delta_int}.  Moreover, $W\neq 0$ ensures
that they do indeed form a fundamental set of solutions of the
envelope equation; a rigorous proof is given in
Appendix~\ref{app:proof}, thereby justifying that
Eq.~(\ref{eq:rho=abc-lin-combi}) represents the general solution of
Eq.~(\ref{eq:rho-linear}).  The linear
decomposition~(\ref{eq:rho=abc-lin-combi}) together with the
constraint~(\ref{eq:abc-constr-k}) play a pivotal role in our work, as
we show next.

\subsection{Matching equations}

\label{sec:match}

Inside the \emph{inner} region ($0<R<R_{\rm in}$) we employ two linear
independent solutions ($\phi,\ \chi$) of the radial
equation~(\ref{eq:radial}), and we ensure $\phi(R)\to0$ when $R\to0$,
such that $\phi$ is the regular solution.  We now use
Eq.~(\ref{eq:rho=abc-lin-combi}) to express the \emph{outer} envelope
$\rho$ in terms of $\phi$ and $\chi$.  We emphasize that the numerical
methods employed for $\phi$, $\chi$ and $\rho$ must ensure their well
defined energy dependence; this will be inherited by the coefficients
$a(E)$, $b(E)$ and $c(E)$, which are obtained from the matching
conditions
\setlength{\jot}{1.5ex}%
\begin{eqnarray}
a\phi^2 +  b\chi^2 + 2 c\phi\chi &=& \rho 
\nonumber\\
a\phi\phi' +  b\chi\chi' +  c\big(\phi\chi\big)'
&=& \textstyle\frac 1 2\rho'
\label{eq:match}
\\
a\Big(\phi'\Big)^2 +b\Big(\chi'\Big)^2 +2c\phi'\chi'
&=& \textstyle \frac 1 2\rho'' -U\!\rho.
\nonumber
\end{eqnarray}
The coefficients $a$, $b$ and $c$ are independent of the matching
point; thus, in principle, the matching conditions could be imposed
anywhere; however, in practice, the matching point should be located
near $R_\text{in}$.  Indeed, the outer phase $\theta$ cannot be
propagated inside the inner well, as explained in
Sec.~\ref{sec:rho}\@.  Conversely, if $\phi$ and $\chi$ were
propagated outside the inner well, they would increase through the
barrier and become linearly dependent.  Hence, as depicted in
Fig.~\ref{fig:Veff}, the most convenient choice for the matching point
is the inner turning point $R_\text{in}$.

The $3\times3$ linear system of equations~(\ref{eq:match}) is
solved in an elementary way; first, we find that the determinant
$\Delta$ is given by a simple
expression, $\Delta=W^3\neq 0$, with $W$ the
(nonvanishing) Wronskian of $\phi$ and $\chi$;  then, the coefficients
$a$, $b$ and $c$ are obtained as the unique solution,
\begin{eqnarray}
 W^2a &=&  \rho\left(\chi'-\frac{\chi\rho'}{2\rho}\right)^2
       +\frac{k^2\chi^2}\rho
\nonumber\\
 W^2b &=&  \rho\left(\phi'-\frac{\phi\rho'}{2\rho}\right)^2
       +\frac{k^2\phi^2}{\rho}
\label{eq:abc=phi-chi}\\
 W^2c &=&  -\rho\left(\phi'-\frac{\phi\rho'}{2\rho}\right)
                       \left(\chi'-\frac{\chi\rho'}{2\rho}\right)
        -\frac{k^2\phi\chi}{\rho},
\nonumber
\end{eqnarray}
with $\phi$, $\chi$ and $\rho$ evaluated at the matching point.
The coefficients $a$, $b$ and $c$ can now be used to obtain the
phase shift.

\subsection{Extracting the scattering phase shift}

\label{sec:delta}

According to Eq.~(\ref{eq:delta=-theta}), in order to find the phase
shift, we need to extend the outer phase into the inner region; this
can be accomplished using
Eqs.~(\ref{eq:theta-prime-rho})--(\ref{eq:abc-constr-k}), as shown in
Appendix~\ref{sec:extend}\@.  The key result is
Eq.~(\ref{eq:theta-at-R}), which yields the outer phase at $R=0$.  For
the sake of clarity, we set $W=k$ in Eq.~(\ref{eq:theta-at-R}) to
simplify the expression of the outer phase,
\begin{equation*}
\theta(0) =  \theta_* - \pi N_* -\alpha_* +\arctan(c),
\end{equation*}
where $\theta_*\equiv\theta(R_\text{in})$ stems from the outer-region
propagation,  $N_*$ is the number of nodes of $\chi$ in the inner
region, and
\begin{equation}
  \label{eq:alpha}
\alpha_*  = \arctan(c+a z_*),
\end{equation}
with $z_*=\frac{\phi(R_\text{in})}{\chi(R_\text{in})}$.  Finally, we
substitute $\theta(0)$ in Eq.~(\ref{eq:delta=-theta}) to find
\begin{equation}
\label{eq:delta=simple}
\delta_\ell = \ell\frac\pi2 -\theta_*
  +\pi N_* +\alpha_* +\arctan(-c).
\end{equation}
The phase shift is thus expressed in terms of the coefficients $a$ and
$c$ that we obtained in the previous section.  The last term in the
equation above, namely $\arctan[-c(E)]$, yields the width $\Gamma$ for
ultra-narrow resonances, as we shall see in Sec.~\ref{sec:Gamma}\@.
However, in preparation for extracting $\Gamma$, we first employ a
phase-amplitude parametrization for the inner solutions $\phi$ and
$\chi$ in the next section, which yields simpler expressions for the
coefficients $a$, $b$ and $c$.

\subsection{Locally adapted solutions in the inner region}

\label{sec:opt}

Although $\phi$ and $\chi$ can be obtained as numerical solutions of
the radial equation, we prefer instead to employ the phase-amplitude
method in the inner region (similar to the outer region).  This will
make it possible to express the coefficients $a,\ b$ and $c$ in terms
of an inner-region phase which has a smooth energy dependence.

Let $\varrho$ denote the envelope inside the inner region, and $\beta$
the corresponding phase function,
\begin{equation}\label{eq:beta}
\beta(R) \equiv  \! \int_0^R \!\! \frac q {\varrho(r)} dr,
\end{equation}
where the parameter $q>0$ can be chosen conveniently.  We emphasize
that the inner and outer envelope functions ($\varrho$ and $\rho$,
respectively) are \emph{different} solutions on the envelope equation;
consequently, the phase functions $\beta$ and $\theta$ differ
nontrivially.  A simple optimization procedure \cite{arxiv:milne-opt}
is employed in the inner region to ensure the smoothness of $\varrho$
and $\beta$, which we now use to construct $\phi$ and $\chi$,
\begin{equation}\label{eq:phi-chi}
  \phi = \sqrt{\varrho}\sin\beta, \qquad
  \chi = \sqrt{\varrho}\cos\beta.
\end{equation}
We remark that Eq.~(\ref{eq:beta}) ensures $\beta=0$ at $R=0$.  Thus,
$\phi$ is the regular solution, as desired; moreover,
Eqs.~(\ref{eq:beta}) and (\ref{eq:phi-chi}) yield the Wronskian
$W=\phi'\chi-\phi\chi'=q$.  We now substitute Eqs.~(\ref{eq:beta}) and
(\ref{eq:phi-chi}) in Eq.~(\ref{eq:abc=phi-chi}) to rewrite the
coefficients $a,\ b$ and $c$ in terms of the inner phase $\beta$,
\begin{eqnarray}
 a  &\;=\;&  u\,\cos^2(\beta+\eta) +\varepsilon\cos^2\beta
\nonumber\\
 b  &\;=\;&  u\,\sin^2(\beta+\eta) +\varepsilon\sin^2\beta
\label{eq:abc-raw}\\
\nonumber
 c  &\;=\;& - u\,\sin(\beta+\eta)\cos(\beta+\eta)
 -\varepsilon\sin\beta\cos\beta.
\end{eqnarray}
In the equations above and hereafter, $\beta=\beta(R_\text{in})$.  The
inner and outer envelopes (and their derivatives) at the matching
point also appear in Eq.~(\ref{eq:abc-raw}) via the quantities $\eta$,
$u$ and $\varepsilon$,
\begin{equation}
    \label{eq:cot-eta}
\cot\eta = \frac{\varrho}{2q}
               \left(\frac{\varrho'}{\varrho}
                    -\frac{\rho'}\rho\right),
\end{equation}
\begin{equation}
  \label{eq:u-and-eps}
u = \frac\rho{\varrho} \csc^2\eta,
\qquad
\varepsilon =  \frac{\varrho} \rho \left(\frac k q\right)^2.
\end{equation}
The three parameters above are interrelated, as they obey the
relationship $u\varepsilon = \Big(\frac{k}{q}\csc\eta\Big)^2$.

The equations above render the phase shift $\delta_\ell$ in
Eq.~(\ref{eq:delta=simple}) expressed exclusively in terms of
quantities obtained from the phase-amplitude formalism; indeed,
  $N_*=\nint[\beta(R_\text{in})/\pi]$, where
$\nint[\cdots]$ stands for  nearest integer, while making use of
$z_*=\frac{\phi(R_\text{in})}{\chi(R_\text{in})}=\tan\beta(R_\text{in})$,
$\alpha_*$ in Eq.~(\ref{eq:alpha}) reads
\begin{eqnarray}
\alpha_*  &=&   \arctan(c+a\tan\beta)
     \label{eq:alpha-star}
  \\
      \nonumber
  &=& \arctan\left[-u\sin(\eta)\cos(\beta+\eta)\sec(\beta)\right]
  \\
  &=&
  -\arctan\left(\frac{\rho\cos(\beta+\eta)}{\varrho\sin\eta\cos\beta}\right).
  \nonumber
\end{eqnarray}

Finally, we remark that the equations in this section remain valid if
the inner envelope $\varrho$ has residual oscillations; thus, strictly
speaking, the inner envelope $\varrho$ need not be smooth.  However,
the optimization method~\cite{arxiv:milne-opt} that we devised for
honing in on the smooth envelope is advantageous in practice, provided
that a well defined $E$-dependence for $\varrho$ is ensured; indeed,
attention must be paid when employing optimization, as the inner
envelope will be initialized with values which are \emph{numerical}
functions of energy.

\section{theory: envelope rescaling and resonance widths}
\label{sec:theory-2}

\subsection{Envelope rescaling}
\label{sec:scaling}

As explained in Sec.~\ref{sec:rho}, the outer envelope follows a
quasi-exponential behavior under the barrier when $E<E_\text{top}$,
which yields $\rho(R_\text{in})\ggg 1$.  Hence, the coefficients $a$,
$b$ and $c$ can reach exceedingly large values and have to be
rescaled; indeed, $\rho(R)$ is rescaled during its propagation through
the barrier, in order to avoid numerical overflow.  Therefore, at the
end of the propagation, the value of $\rho(R_\text{in})$, and thus $u$
and $\varepsilon$, will be represented logarithmically.

We remark that, although $a$, $b$ and $c$ are independent of the
matching point, the parameters $\eta$, $u$ and $\varepsilon$ do depend
on its location.  Hence, if $u$ (or $\rho$ itself) were used as
scaling factor, the rescaled coefficients would depend on the matching
point.  Although this would not entail any difficulty, it is possible
to rescale the coefficients such that they do remain formally
independent of the matching point.  Namely, we choose the quantity
$\upsilon\equiv u+\varepsilon$ as the scaling factor; from
Eq.~(\ref{eq:abc-raw}) we find
\begin{equation}
  \label{eq:scaling-factor}
\upsilon = u+\varepsilon = a+b,
\end{equation}
which is  independent of the matching point, and we define the
scaled coefficients according to
\begin{equation}
    \label{eq:scaling}
     \tilde a \equiv \frac a{\upsilon},
\qquad\tilde b \equiv \frac b{\upsilon},
\qquad\tilde c \equiv \frac c{\upsilon}.
\end{equation}
Equation~(\ref{eq:abc-raw}) can now be recast as
\begin{eqnarray}
\tilde a  &\;=\;& \tilde u \cos^2(\beta+\eta) +\tilde\varepsilon\cos^2\beta
\nonumber\\
\tilde b  &\;=\;& \tilde u \sin^2(\beta+\eta) +\tilde\varepsilon\sin^2\beta
\label{eq:abc-tilde}\\
\nonumber
\tilde c  &\;=\;& -\tilde u \sin(\beta+\eta)\cos(\beta+\eta)
 -\tilde\varepsilon\sin\beta\cos\beta,
\end{eqnarray}
where the scaled parameters
\[
\tilde u \equiv \frac u{\upsilon}, \qquad
\tilde\varepsilon \equiv \frac\varepsilon{\upsilon}
\]
obey the simple relationship
\[
\tilde u + \tilde\varepsilon = 1,
\]
which render the scaled coefficients of the order of unity.

\subsection{Ultra-narrow resonances}
\label{sec:bg+res}

For scattering energies $E$ sufficiently lower than $E_\text{top}$, we
enter the regime of ultra-narrow resonances, characterized by
$\tilde\varepsilon\sim\rho^{-2}(R_\text{in})\lll\tilde{u}\approx1$.
This simplifies greatly the expressions of the scaled coefficients;
indeed, Eq.~(\ref{eq:abc-tilde}) becomes
\begin{eqnarray}
  \tilde a(E)  &\;\approx\;& \cos^2\beta_\text{full}(E)
  \nonumber
\\
\tilde b(E)  &\;\approx\;&  \sin^2\beta_\text{full}(E)
\label{eq:abc-clean}
\\
\tilde c(E)  &\;\approx\;& -\sin\beta_\text{full}(E)\cos\beta_\text{full}(E),
\nonumber
\end{eqnarray}
where the phase
\begin{equation}
  \label{eq:beta-full}
\beta_\text{full} \equiv \beta+\eta
\end{equation}
represents the full  contribution from the inner region
\emph{and} the barrier; see Appendix~\ref{app:thor}.

Ultra-narrow resonances correspond to metastable (quasibound) states,
and their positions ($E_\text{res}$) can be obtained as the roots of
$\beta_\text{full}(E)=N\pi$ with $N$ a positive integer.  Hence, the
resonance positions are the minima of $\tilde b(E)$, i.e., the roots of
$\sin\beta_\text{full}=0$.  Note that we also have $\tilde c(E)=0$ at
$E=E_\text{res}$.  We remark that methods which are suitable for bound
states can be used to find the positions $E_\text{res}$ of quasi-bound
states.  On the other hand, the vanishingly small widths ($\Gamma$) of
such resonances are difficult to obtain.

In preparation for the next section, where the resonance width
$\Gamma$ will be extracted, we first rewrite $\delta_\ell$ in
Eq.~(\ref{eq:delta=simple}) as a sum of background and resonant
contributions, and we analyze the resonant phase shift in detail.  For
scattering energies sufficiently lower than $E_\text{top}$, the large
barrier plays the role of a repulsive wall.  Therefore, the inner
region is inaccessible (unless $E\approx E_\text{res}$) and we
identify the background term,
\begin{equation}\label{eq:delta-bg}
\delta_\ell^{\rm bg}(E) \equiv \ell\frac\pi2 - \theta_*(E),
\end{equation}
which is given by the outer phase $\theta_*(E)=\theta(E;R_\text{in})$,
with $R_\text{in}$ playing the same role as $R=0$ in
Eq.~(\ref{eq:delta=-theta}).  The remaining terms in
Eq.~(\ref{eq:delta=simple}) give the contribution of the inner region,
which we interpret as the resonant part of the phase shift,
\begin{equation}
  \delta_\ell^{\rm res}(E) \equiv \pi N_*(E)
  + \alpha_*(E)  \arctan[-c(E)].
\label{eq:delta-res}
\end{equation}
To simplify our notation, we shall omit the subscript $\ell$ for the
remainder of this article, and we rewrite
Eq.~(\ref{eq:delta=simple}) as
\[
\delta(E) = \delta^{\rm bg}(E) + \delta^{\rm res}(E).
\]
As we explain next, the resonant phase shift is very nearly constant
between resonances, $\delta^\text{res}(E)\approx N\pi$.  Thus, we have
\begin{equation}\label{eq:delta=bg-mod-pi}
  \delta(E)
  \stackrel{{\scriptscriptstyle\text{mod}\,\pi}}=
  \delta^\text{bg}(E),
\qquad E\neq E_\text{res},
\end{equation}
which confirms the interpretation of $\delta^\text{bg}$ in
Eq.~(\ref{eq:delta-bg}) as the background phase shift.

In order to understand the energy dependence of $\delta^{\rm res}(E)$,
we first recall that $N_*(E)$ is an integer-valued step function;
secondly, in the regime of ultra-narrow resonances we have
$\alpha_*(E)\approx\arctan(\pm\infty)=\pm\frac\pi2$, due to
$u\sim\rho\to\infty$ in Eq.~(\ref{eq:alpha-star}).  Similarly,
$\arctan[c(E)]\approx\arctan(\pm\infty)=\pm\frac\pi2$, and thus the
last two terms in Eq.~(\ref{eq:delta-res}) yield
$\alpha_*-\arctan(c)\approx\pm\pi$ or zero.  Consequently,
$\delta^{\rm res}(E)$ is to an excellent approximation a piecewise
constant function, whose values are integer multiples of $\pi$.  More
precisely, $\delta^{\rm res}(E)$ follows a stepwise behavior,
increasing sharply by $\pi$ at each resonance, as we explain next.

The behavior of $\delta^{\rm res}(E)$ can be fully elucidated by a
more detailed analysis of the terms in
Eq.~(\ref{eq:delta-res}). First, the discontinuous steps of
$N_*(E)=\nint[\beta/\pi]$ when
$\beta\stackrel{{\scriptscriptstyle\text{mod}\,\pi}}=\frac\pi2$ are
irrelevant, as each step ($+\pi$) due to $\pi N_*(E)$ is canceled by
an opposite ($-\pi$) step given by
$\alpha_*(E)=\arctan(c+a\tan\beta)$, due to $\tan\beta$ jumping from
$+\infty$ to $-\infty$.  Second, we observe that both $\tilde{a}(E)$
and $\tilde{c}(E)$ will vanish when $\cos\beta_\text{full}=0$; see
Eq.~(\ref{eq:abc-clean}).  The roots of $\cos\beta_\text{full}=0$ are
interspersed between the roots of $\sin\beta_\text{full}=0$, i.e, the
zeros of $b(E)$.  The latter give the resonance positions
$E_\text{res}$, while the common zeros of $a(E)$ and $c(E)$ are
completely unremarkable despite the fact that both $\alpha_*(E)$ and
$\arctan[c(E)]$ in Eq.~(\ref{eq:delta-res}) vary rapidly in their
vicinity; indeed, using the definition~(\ref{eq:alpha}) of $\alpha_*$
and the constraint $1+c^2=ab$ (see Eq.~(\ref{eq:abc-constr-k}) with
$W=k$), we find that the last two terms in Eq.~(\ref{eq:delta-res})
cancel nearly perfectly,
\begin{figure}[b]
\includegraphics[width=0.95\linewidth]{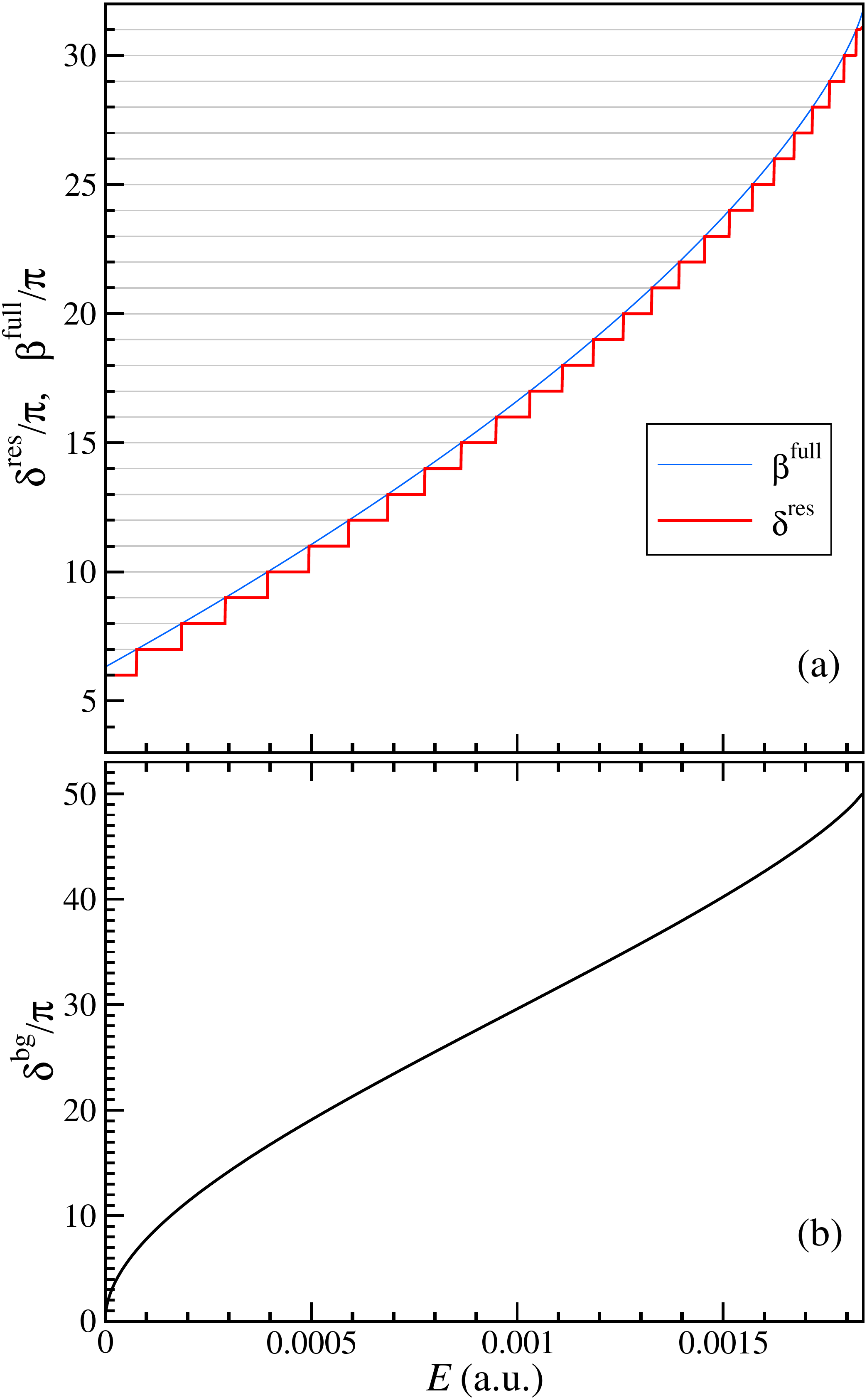}
\caption{\label{fig:delta-res} Energy dependence of the resonance
  phase shift $\delta^\text{res}$ (thick red line) and
  $\beta_\text{full}=\beta+\eta$ (thin blue line) in the top panel
  (a), and background phase shift $\delta^\text{bg}$ in the bottom
  panel (b).  The full phase shift
  $\delta=\delta^\text{bg}+\delta^\text{res}$ is shown in
  Fig.~\ref{fig:abc}(c), where the resonance positions are indicated.
  The results were obtained using the potential
  energy~(\ref{eq:Vpot}).}
\end{figure}
\begin{eqnarray*}
\alpha_*  &-& \arctan(c) =  \arctan(c+az_*) - \arctan(c)
\\
          &=& \arctan\left(\frac{az_*}{1+c^2+acz_*}\right)
 = \arctan\left(\frac{z_*}{b+cz_*}\right) \approx 0.
\end{eqnarray*}
This expression vanishes because $b\approx\infty$ when
$\sin\beta_\text{full}\not\approx0$.
Finally, one is left with the only possible explanation for the
stepwise behavior of $\delta^{\rm res}(E)$.  Namely, it
stems solely from the last term in Eq.~(\ref{eq:delta-res}),
\[
\arctan[-c(E)] = -\arctan\big[\upsilon(E)\tilde c(E)\big].
\]
Indeed, at $E_\text{res}$ we have $c=0$ (and $b\approx 0$) due to
$\sin\beta_\text{full}=0$.  Moreover, the
derivative $\dot c \equiv\frac{dc}{dE}$ is very large at $E=E_\text{res}$,
\begin{equation}
\label{eq:c-dot}
\dot c(E_\text{res})
= \upsilon(E_\text{res})\dot{\tilde c}(E_\text{res})
 = -\upsilon(E_\text{res})\dot\beta_\text{full}(E_\text{res}).
\end{equation}
Hence, as $E$ increases within a narrow window around $E_\text{res}$,
$c(E)$ decreases rapidly (practically from $+\infty$ to $-\infty$),
which yields a rapid \emph{increase} of $\arctan[-c(E)]$ from
$-\frac\pi2$ to $+\frac\pi2$.  This is in agreement with the well
known signature of scattering resonances; namely, the increase by
$\pi$ of the phase shift at each resonance, as depicted in
Fig.~\ref{fig:delta-res}(a).

\begin{figure*}
\includegraphics[height=0.60\textheight,width=\textwidth]{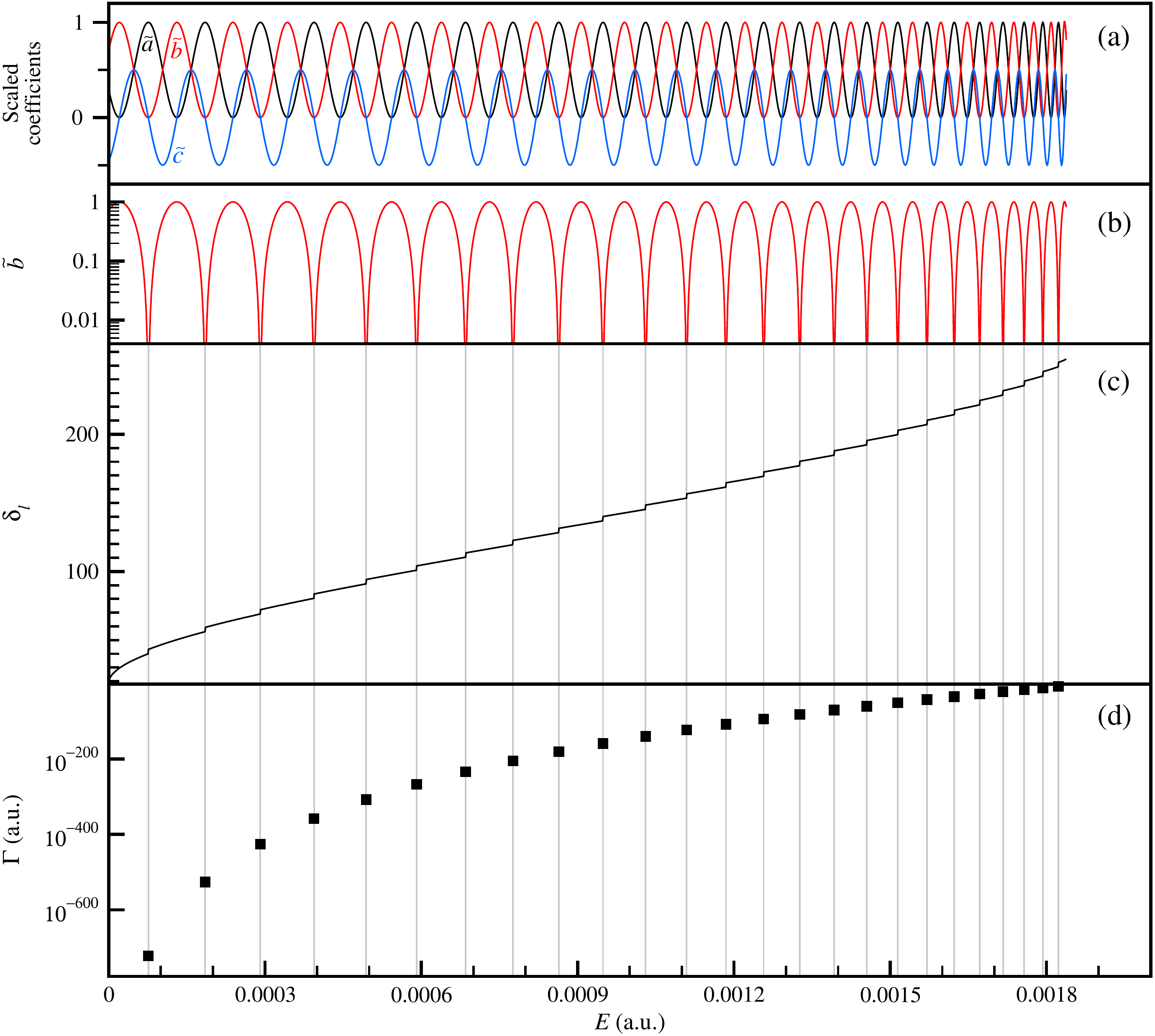}
\caption{\label{fig:abc} Energy dependence of the scaled coefficients
  (a), semilog plot of $\tilde b$ (b), phase shift $\delta_\ell$ (c),
  and resonance widths (d).  The vertical lines mark the positions of
  the resonances.  }
\end{figure*}

\subsection{Resonance widths}
\label{sec:Gamma}

We now extract the widths of ultra-narrow resonances, while the case
of broad resonances (e.g., above-barrier resonances) will be discussed
in Sec.~\ref{sec:above}.  As is well known, the scattering phase shift
$\delta(E)$ increases rapidly when $E\approx E_\text{res}$, and its
derivative $\dot\delta(E)\equiv\frac{d\delta}{dE}$ has a sharp maximum
at $E_{\rm res}$.  We thus analyze $\dot\delta$ to extract the
resonance width $\Gamma$.  For ultra-narrow resonances, the phase
shift can be easily separated into background and resonant
contributions, as shown in the previous section.  Moreover, in the
immediate vicinity of a narrow resonance, the background term is
nearly constant and we neglect it.  Therefore, we need only consider
the resonant phase shift in Eq.~(\ref{eq:delta-res}). 
Specifically, its derivative reads
\begin{equation}\label{eq:dot-delta-res}
 \dot\delta^\text{res}\equiv  \frac{d\delta^\text{res}}{dE} 
\approx \frac d {dE}
\big[\!-\!\arctan c(E)\big]
 =  -\frac{\dot c(E)}{1+c^2(E)},
\end{equation}
where we used  $\dot N_*=0$ and $\dot\alpha_*\approx 0$.
In order to extract the resonance width $\Gamma$, we  employ  the
linear approximation
\begin{equation}\label{eq:c-lin}
  c(E)\approx\dot c_{\rm res}(E-E_{\rm res}),
\end{equation}
with $\dot c_\text{res}\equiv \dot c(E_\text{res})$.  The
linearization~(\ref{eq:c-lin}) is essentially exact within a
sufficiently narrow window $\Delta E$; at the same time, the strong
inequality $\Delta E\ggg\Gamma$ also holds.  Thus, the line shapes of
ultra-narrow resonances are accurately given by
\begin{equation} \label{eq:dot-c-res}
  \dot\delta^\text{res}(E)
  \approx -\frac{\dot{c}_\text{res}}
{1+\big(\dot{c}_\text{res}\big)^2
  \big(E-E_\text{res}\big)^2}.
\end{equation}
Comparing this result to the familiar Breit--Wigner expression,
\begin{equation}\label{eq:BW}
\dot\delta_\text{BW}(E)
= \frac{\frac\Gamma 2}{\big(E-E_\text{res}\big)^2
  + \left(\frac\Gamma 2\right)^2},
\end{equation}
we  identify the resonance width,
\begin{equation}\label{eq:Gamma}
  \frac 2 \Gamma = -\dot c_\text{res}.
\end{equation}
Making use of Eq.~(\ref{eq:c-dot}), we can express the resonance width
in terms of the scaled coefficients,
\[
\frac 2 \Gamma = -\upsilon_\text{res}\dot{\tilde c}_\text{res}
 =  \upsilon_\text{res} \dot\beta^\text{full}_\text{res},
\]
with $\dot{\tilde{c}}_\text{res}\equiv\dot{\tilde{c}}(E_\text{res})$
and $\upsilon_\text{res}\equiv \upsilon(E_\text{res})$.  We emphasize
that the vanishingly small value of $\Gamma$ for ultra-narrow
resonances stems from the huge value of the scaling factor
$\upsilon\approx u$, which in turn is due to the exponential increase
of the envelope through the barrier.  Finally, we remark that the
linearization~(\ref{eq:c-lin}) was used only within a very narrow
window $\Delta E$ around $E_\text{res}$ to facilitate the formal
comparison of the Breit--Wigner formula~(\ref{eq:BW}) with
Eq.~(\ref{eq:dot-c-res}).  However, we evaluate the energy derivative
$\dot{\tilde{c}}(E_\text{res})$ using a high order method for
numerical differentiation based on Chebyshev polynomials covering a
wide energy interval.  Thus, in order to attain high accuracy,
we account fully for the nonlinear behavior of $\tilde c(E)$ and
$\beta^\text{full}(E)$.

\section{Results and discussion}

\label{sec:results}

As an illustrative example, we consider the potential energy employed
in our previous work \cite{delta_int},
\begin{equation}\label{eq:Vpot}
  V(R) = C_{\rm wall}  \,e^{-\frac R{R_\text{wall}}}
\;-\; \frac{C_3}{R^3+R_{\rm core}^3},
\end{equation}
with $C_{\rm wall}=10$, $R_{\rm wall}=1$, $R_{\rm core}=5$ and
$C_3=18$ (all in atomic units), and the reduced mass $\mu=\frac m 2$,
where $m$ is the mass of $^{88}$Sr.  Although $V(R)$ is barrierless,
the effective potential, $V_\text{eff}=V+\frac{\ell(\ell+1)}{2\mu
  R^2}$, will have a centrifugal barrier for $0<\ell\lessapprox 557$.
We are interested in the case of a large barrier, and thus a
sufficiently high value for $\ell$ will be used; namely, $\ell=500$.
As depicted in Fig.~\ref{fig:Veff}, the effective potential
has a large centrifugal barrier and a sufficiently deep well at short
range holding a large number of shape resonances.  Hence, our example
is a suitable representative for potentials which
posses ultra-narrow shape resonances.

Figure~\ref{fig:delta-res} shows the energy dependence of $\delta^{\rm
  bg}(E)$ and $\delta^{\rm res}(E)$, as well as
$\beta_\text{full}(E)$.  It is readily apparent in
Fig.~\ref{fig:delta-res}(a) that the resonant phase shift is constant
between resonances, $\delta^\text{res}(E)\approx N\pi$, with the
integer $N$ increasing by unity for each resonance, as we discussed
in Sec.~\ref{sec:bg+res}\@.

The phase $\beta_\text{full}(E)$ has a smooth energy dependence, as
shown in Figure~\ref{fig:delta-res}(a), which explains the simple
oscillatory behavior of the scaled coefficients in
Fig.~\ref{fig:abc}(a).  The unscaled coefficients follow the same
oscillatory behavior, albeit modulated by the strongly varying
amplitude $\upsilon(E)\approx u(E)$, which is dominated by the
quasi-exponential behavior of the outer envelope
$\rho(E;R_\text{in})$.  However, the scaling~(\ref{eq:scaling}) was
not introduced to merely simplify the plot in Fig.~\ref{fig:abc}(a).
Indeed, the scaling factor $\upsilon(E)$ and the scaled coefficient
$\tilde c (E)$ proved instrumental in extracting the resonance width
$\Gamma$, as discussed in Sec.~\ref{sec:Gamma}.

A semilog plot of $\tilde b(E)$ is shown in Fig.~\ref{fig:abc}(b),
while the phase shift is depicted in Fig.~\ref{fig:abc}(c).  As
discussed in Sec.~\ref{sec:bg+res}, the nearly vanishing minima of
$\tilde{b}(E)$ and hence of $b(E)$ signify resonances, whose positions
are marked by the vertical lines; the widths $\Gamma$ are plotted in the
bottom panel (d).

\begin{figure}[b]
\includegraphics[width=0.94\linewidth]{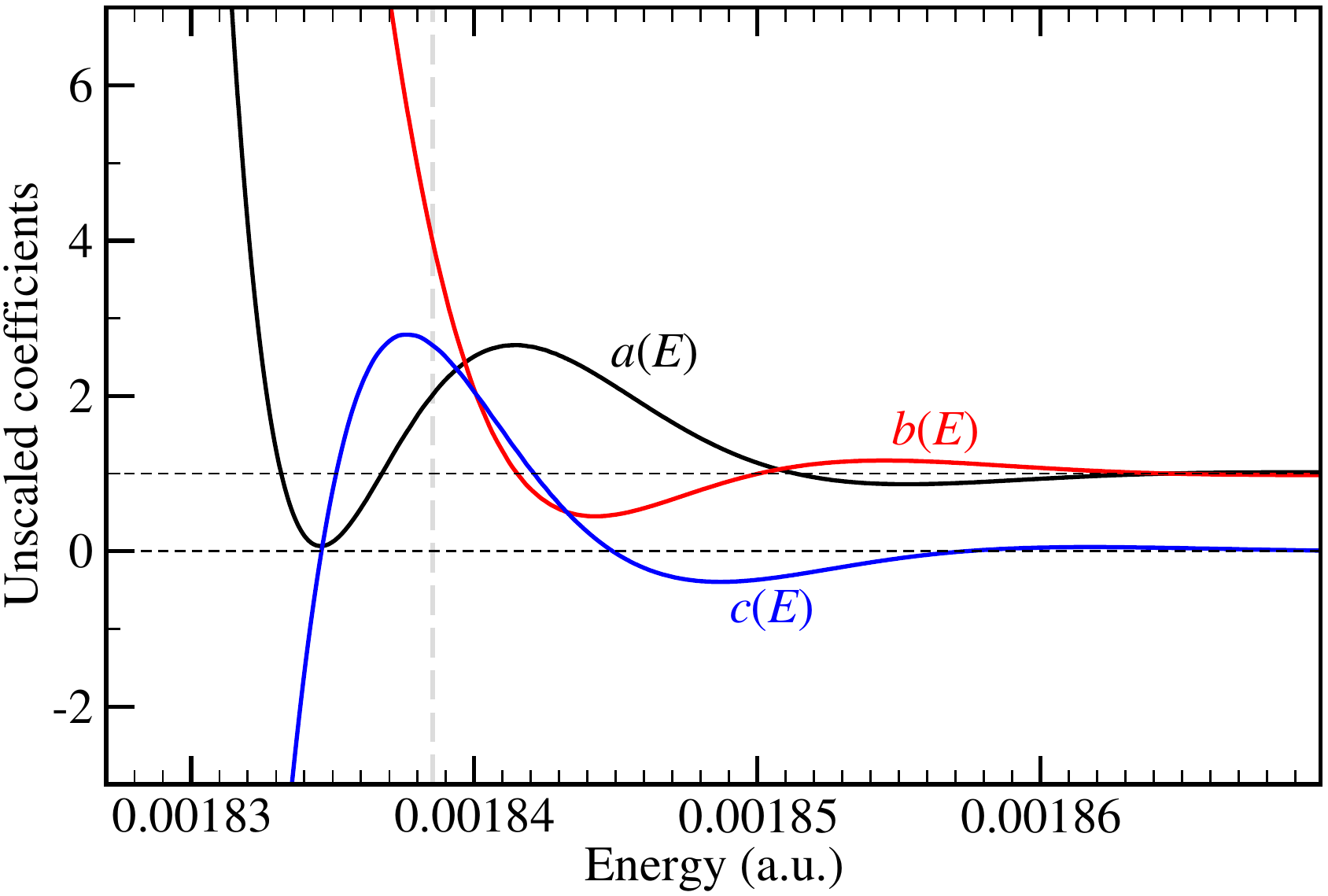}
\caption{\label{fig:true-abc}  Energy dependence of
  the coefficients $a(E)$, $b(E)$ and $c(E)$ in the vicinity of
  $E_\text{top}$, which is marked by the vertical dashed line.  As the
  energy $E$ increase above  the barrier, the coefficients
  $a(E)$ and $b(E)$ converge to unity, while $c(E)$ converges to zero;
  these limits are marked by horizontal dashed lines.  }
\end{figure}

\subsection{Above-barrier resonances}

\label{sec:above}

For scattering energies \emph{just above} the barrier, the situation
is similar to the case $E<E_\text{top}$; namely, a globally smooth
envelope does not exist ($\varrho\neq\rho$).  Hence, it is again
advantageous to combine locally adapted solutions for the inner and
outer regions.  However, global smoothness will be recovered very
quickly when the energy increases above the barrier; this is apparent
in Fig.~\ref{fig:true-abc}, which shows the behavior of the
\emph{unscaled} coefficients ($a$, $b$, $c$) for energies below and
above $E_\text{top}$.  The limits $a(E)\approx b(E)\to1$ and
$c(E)\to0$, which correspond to the globally smooth envelope
$\varrho=\rho$, are eventually attained at high energies.  We remark
that the nonexistence of a globally smooth envelope
($\varrho\neq\rho$) for energies just above the barrier is closely
related to quantum reflection \cite{qr, robin_harald_step,
  Robin_qr_mirror,Robin_qr_mirror2}, which only vanishes at energies
sufficiently high above the barrier (when a globally smooth envelope
does exist).

As is well known, shape resonances may  occur for energies just
above the barrier.  Such  resonances are rather
broad,  and  are in stark contrast with the ultra-narrow
resonances described in the previous section.  We now discuss briefly
an example of a broad above-barrier resonance, which will  shed more
light on ultra-narrow resonances.

\begin{figure}[b]
  \includegraphics[width=0.84\linewidth]{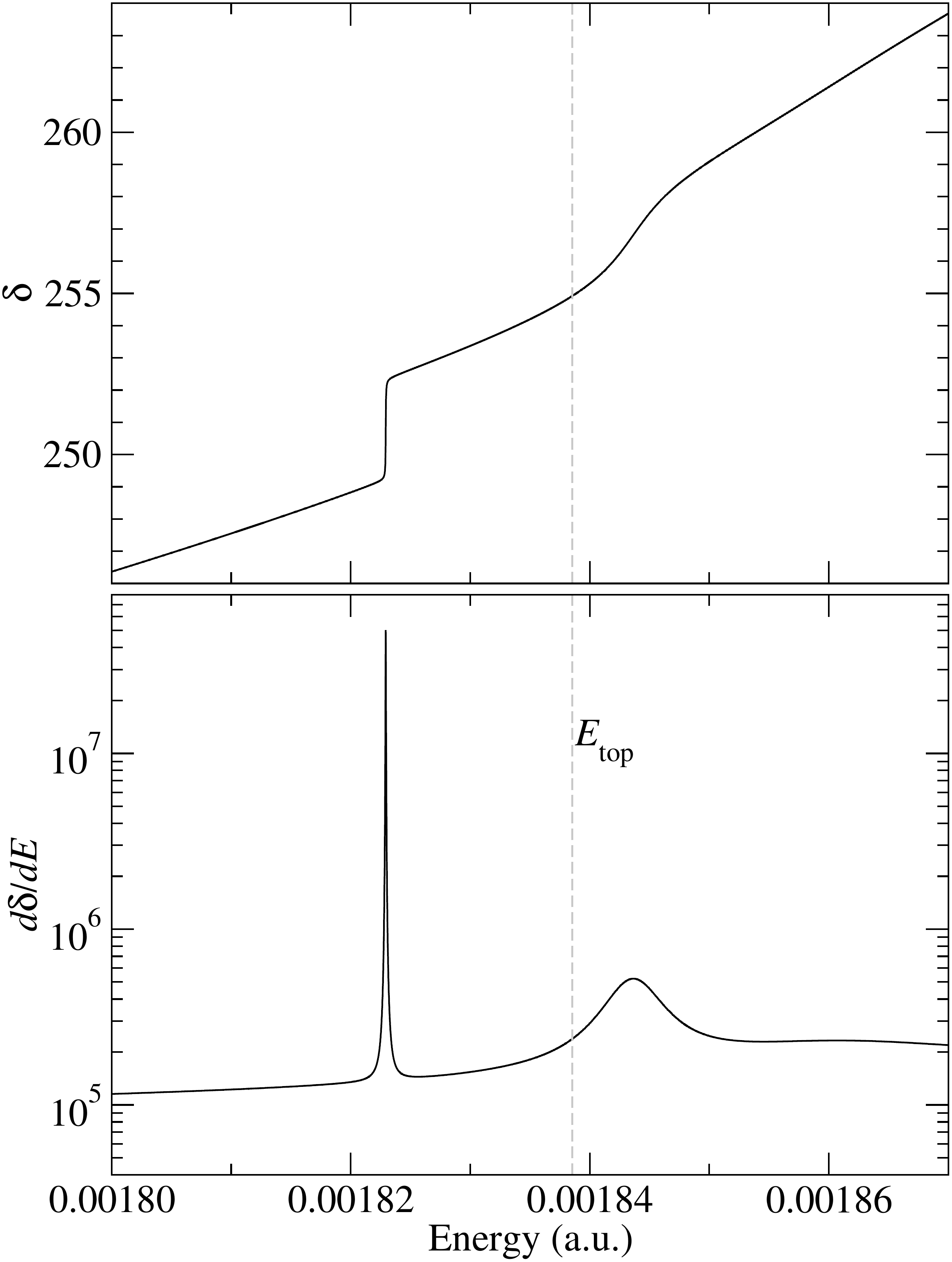}
\caption{\label{fig:top-2-res} Energy dependence of the phase shift
  $\delta$ and its derivative $\dot\delta$ for scattering energies
  near the top of the barrier $E_\text{top}$ (marked by a vertical
  dashed line).  The sharp feature is the first resonance below the
  top of the barrier.  The broad resonance above the barrier is also
  visible (especially in the lower panel).}
\end{figure}

For $E>E_\text{top}$, a convenient choice for the matching point is
$R_\text{top}$ (the location of the barrier top).  However, as the
energy increases above the barrier, the boundary between the inner and
outer regions becomes arbitrary; hence, the interpretation of the
outer-region contribution~(\ref{eq:delta-bg}) as the background phase
shift loses its meaning.  Thus, although our method is still useful
for computing the phase shift, the extraction of the resonance width
(and position) must be performed by \emph{fitting} the resonance
line-shape using the  Breit--Wigner formula.  The fitting
procedure must also include an energy dependent background, as it
cannot be neglected in this case; indeed, for broad resonances,
$\delta^\text{bg}(E)$ may vary significantly within $\Delta
E\sim\Gamma$.

Figure~\ref{fig:top-2-res} shows the behavior of the phase shift
$\delta(E)$ and its derivative $\dot\delta(E)$ for energies $E$ near
the top of the barrier.  Two resonances are readily apparent; namely,
the first resonance under the barrier, which is sufficiently narrow to
be analyzed as explained in Sec.~\ref{sec:Gamma}, and a broad
resonance above the barrier.  The latter is resolved by fitting its
lineshape, as mentioned above.  Specifically, we employ
\[
\dot\delta_\text{fit}(E)
 = \dot\delta_\text{BW}(E) +\dot\delta^\text{bg}_\text{fit}(E),
\]
with $\dot\delta_\text{BW}(E)$ given in Eq.~(\ref{eq:BW}) and a low
degree polynomial for the background term
$\dot\delta^\text{bg}_\text{fit}(E)$ to extract the resonance position
$E_{\rm res} = 1.84362\times 10^{-3}\text{ a.u.}$ and width $\Gamma =
5.628\times 10^{-6}\text{ a.u.}$

We emphasize that the fitting procedure can only be used when
resonances are sufficiently broad for their lineshapes to be resolved
via a numerical scan of $E$ within $\Delta E\sim\Gamma$.  Although
this is a trivial observation, we need to bring it to the fore,
because the direct fitting method cannot be used when $\Gamma$ is
vanishingly small.  Indeed, scanning through a narrow energy window
$\Delta E\sim\Gamma$ in the vicinity of $E=E_\text{res}$ cannot be
done if $\log_{10}(E_\text{res}/\Gamma) > N_\text{digits}$, where
$N_\text{digits}$ is the number of digits available in machine
arithmetic.  This simple limitation of computer arithmetic is a severe
obstacle for directly resolving ultra-narrow resonances, but is almost
never mentioned in the literature; a notable exception is
Ref.~\cite{sidky-barrier}.

\subsection{Accuracy test}

In order to explore the numerical accuracy of our method, we use
 the following result from Breit and Wigner \cite{breit-wigner},
\begin{equation}
  \label{eq:bw-int}
k_{\rm res} \!\! \int_0^{R_{\rm out}} \!\!\!\!\! \left|\psi_{\rm res}(R)\right|^2dR
\approx 2\frac{E_\text{res}}\Gamma,
\end{equation}
which was employed in similar work on resonances~\cite{allison-1969,
  jackson-wyatt-1970, sando-dalgarno-1971,res-jim}.  In the equation
above, $R_\text{out}$ is the outermost turning point and
$\psi_\text{res}(R)\equiv\psi(E_\text{res};R)$ is the physical wave
function normalized to unit amplitude asymptotically, i.e.,
$\psi\sim\sin\big(kR-\ell\frac\pi2+\delta\big)$, which we now express
in terms of the regular solution $\phi$ and the Jost function
$\mathscr F$ \cite{newton,taylor},
\begin{equation}
  \label{eq:psi=phi-over-F}
\psi(E;R) = \frac{\phi(E;R)}{\mathscr F(E)}.
\end{equation}
The regular solution has the asymptotic behavior
\[
\phi(R) \ \xrightarrow{R\to\infty}\ \,
       \mathscr A\sin\big(kR-\ell\textstyle\frac\pi2\big)
      +\mathscr B\cos\big(kR-\ell\textstyle\frac\pi2\big),
\]
with $\frac{\mathscr B}{\mathscr A}=\tan\delta$ and
$\mathscr{A}-i\mathscr{B=F}$ the Jost function.  We recall that the
regular solution $\phi$ was employed in the linear
decomposition~(\ref{eq:rho=abc-lin-combi}) of the outer envelope; the
coefficients $a$, $b$ and $c$ in Eq.~(\ref{eq:rho=abc-lin-combi}) obey
the constraint $ab=c^2+1$ (see Eq.~(\ref{eq:abc-constr-k}), with
$W=k$).  Due to the constraint, only $a$ and $c$ appear in the phase
shift expression~(\ref{eq:delta=simple}), while $b$ does not.
However, the coefficient $b$ is directly related to the Jost function;
specifically, it can be shown that
\[
b=\mathscr{A}^2+\mathscr B^2=|\mathscr F|^2.
\]
We now make use of the constraint~(\ref{eq:abc-constr-k}) yet
again to write
\[
\frac 1{|\mathscr F|^2} = \frac 1 b =\frac a {1+c^2},
\]
which we substitute in Eq.~(\ref{eq:psi=phi-over-F}) to obtain
\[
\big|\psi(E;R)\big|^2=\frac{a(E)}{1+c^2(E)} \phi^2(E;R).
\]

\begin{figure}[t]
\includegraphics[width=0.99\linewidth]{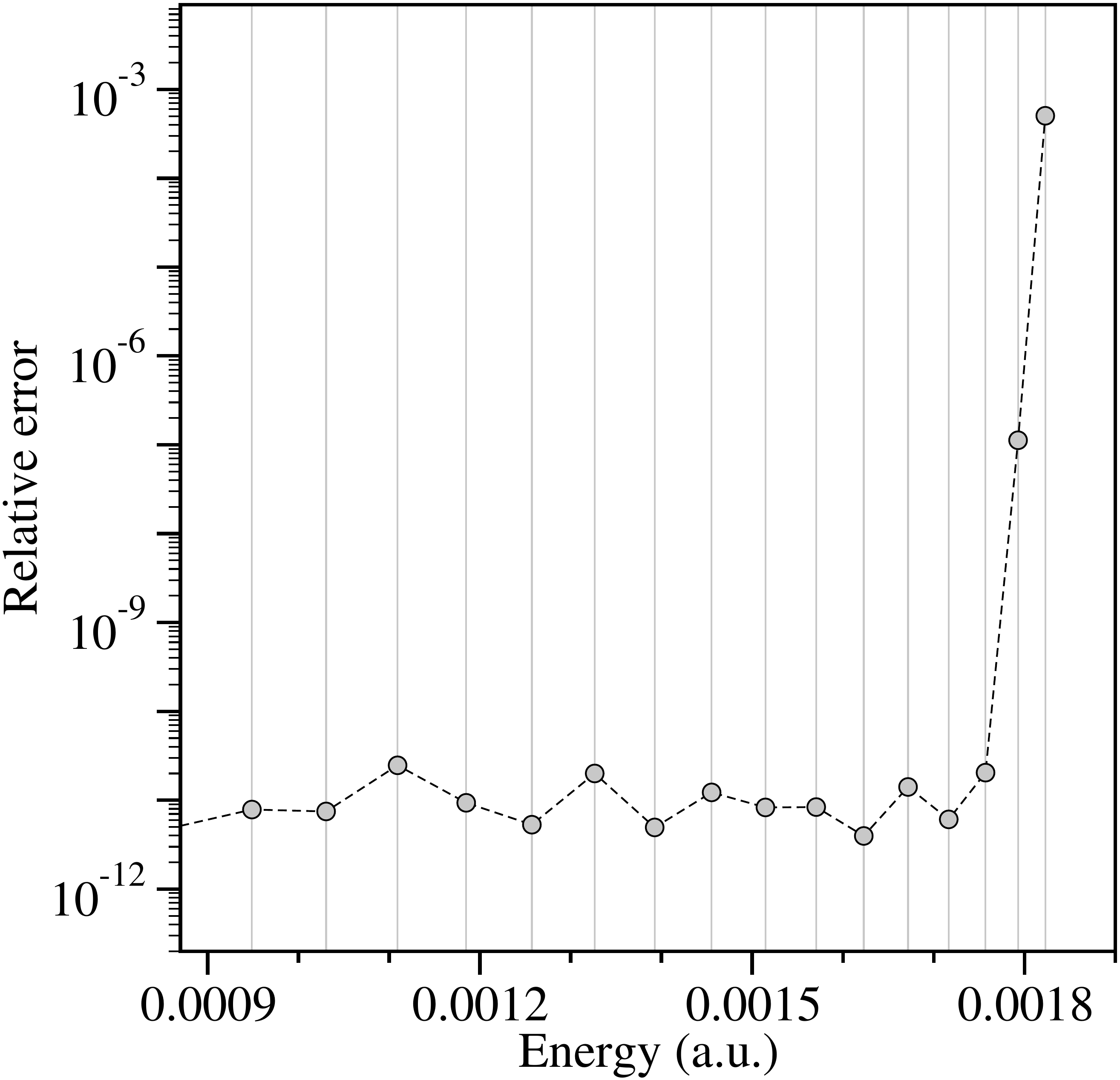}
\caption{\label{fig:jost} The relative error
  $|\mathscr{S-\tilde{S}}|/\mathscr S$, where $\mathscr S$ is the
  integral on the left-hand-side of Eq.~(\ref{eq:test}) and
  $\tilde{\mathscr S}\equiv|\dot{\tilde c}_\text{res}|
  E_\text{res}/k_\text{res}$ is the right-hand-side.
  $\mathscr{\tilde{S}}$ was obtained using the value of $\dot{\tilde
    c}$ evaluated as described in Sec.~\ref{sec:Gamma}, while
  $\mathscr S=\int\!\phi^2$ was computed by numerical quadrature.  }
\end{figure}

For energies within a narrow window centered on $E_\text{res}$, we use
the approximations
\begin{eqnarray*}
  a(E)  &\approx&  a_\text{res}
  \\
  c(E)  &\approx&  \dot{c}_\text{res}(E-E_\text{res})
  \\
\phi(E;R) &\approx& \phi(E_\text{res};R).
\end{eqnarray*}
The latter holds for $R$ throughout the inner region and most of the
barrier, and thus the probability density inside the inner potential
well reads
\[
\big|\psi(E;R)\big|^2 \approx
\frac{a_\text{res}}{1+\dot c^2_\text{res}(E-E_\text{res})^2}
\phi^2_\text{res}(R),
\]
where $\phi_\text{res}(R)\equiv\phi(E_\text{res};R)$.  We now
substitute $|\dot c_\text{res}|=\frac2\Gamma$ from
Eq.~(\ref{eq:Gamma}) and make use of the scaling~(\ref{eq:scaling}) to
recast the expression above such that the familiar Breit--Wigner
expression, i.e., the Lorentzian energy dependence sharply peaked at
$E=E_\text{res}$, is made explicit for the wave function itself,
\[
\big|\psi(E;R)\big|^2 \approx
 \frac{\frac\Gamma 2}{\big(E-E_\text{res}\big)^2
   + \left(\frac\Gamma 2\right)^2} \big|\dot{\tilde c}_\text{res}\big|^{-1}
 \phi^2_\text{res}(R).
\]
For  $E=E_\text{res}$ the equation above reads
\[
\big|\psi_\text{res}(R)\big|^2 \approx\frac 2\Gamma
\big|\dot{\tilde c}_\text{res}\big|^{-1} \phi^2_\text{res}(R),
\]
which we now use to rewrite Eq.~(\ref{eq:bw-int}),
\begin{equation}\label{eq:test}
\int_0^{R_{\rm out}} \!\!\! \phi^2_{\rm res}(R)\,dR
\approx \big|\dot{\tilde c}_\text{res}\big| \frac{E_{\rm res}} {k_{\rm res}}.
 \end{equation}

We emphasize that the approximations used above, as well as in
deriving the Breit--Wigner result~(\ref{eq:bw-int}), are excellent for
ultra-narrow resonances.  Indeed, Fig.~\ref{fig:jost} shows that
Eq.~(\ref{eq:test}) is valid to high accuracy for the resonances
located deep below the top of the barrier, which demonstrates that our
numerical approach can reach a high level of precision.  The
approximate nature of Eq.~(\ref{eq:test}) is only visible for the
highest two resonances located just under the top of the barrier.
Finally, we remark that although Eq.~(\ref{eq:bw-int}) yields
essentially exact results for the widths of ultra-narrow resonances,
the resonantly enhanced amplitude of $\psi_\text{res}(R)$ at short
range cannot be pinned down by scanning the energy directly when
$\Gamma\lll E_\text{res}$ (see discussion at the end of
Sec.~\ref{sec:above}).  Nevertheless, even if the phase-amplitude
approach is not employed, it does suggest a simple remedy for finding
the correct physical wave function $\psi_\text{res}$ when solving the
radial equation~(\ref{eq:radial}) directly;  namely, the resonance
position $E_\text{res}$ is first found, and subsequently the (unknown)
phase shift $\delta$ is varied instead of the energy (which is kept
fixed).  Thus, for $E=E_\text{res}$, the solution $\psi_E$ is
initialized asymptotically using
$\psi_E(\delta;R)=\cos(\delta)j_\ell(kR)+\sin(\delta)n_\ell(kR)
\approx\sin(kR-\ell\frac\pi2+\delta)$, and is propagated inward.
The phase shift $\delta$ is then adjusted to maximize the short-range
amplitude of $\psi_E(\delta;R)$.

\section{Summary and outlook}
\label{sec:end}

The appeal of Milne's phase-amplitude representation \cite{milne}
stems from the fact that it only requires the computation of slowly
varying phase and amplitude functions instead of highly oscillatory
wave functions.  However, this advantage cannot be fully exploited
unless special algorithms are devised for honing in on the smooth
solution.  For scattering problems, an efficient method was developed
by the present authors~\cite{delta_int} for computing the smooth
amplitude in the asymptotic region.  On the other hand, for
classically allowed regions of finite extent, an optimization
procedure is needed to find the smooth amplitude; we have
developed such an optimization algorithm~\cite{arxiv:milne-opt} for
locally adapted solutions, which we employed in this work.

We recently formulated an integral representation for  phase
shifts~\cite{delta_int} based on a phase-amplitude
approach~\cite{milne}; however, our computational method was only
applicable to the case of a single (infinite) classically allowed
region.  In order to generalize our previous work~\cite{delta_int}, we
have now developed a phase-amplitude approach for tackling scattering
potentials with a barrier.  As shown in this article, our new method
is especially useful for energies below the top of the barrier, when
two disjoint classically allowed regions exist.  In particular,
accurate values of resonance widths in the extreme regime of
ultra-narrow resonances ($\Gamma\lll E_\text{res}$) can be easily
obtained.  Numerical results are presented for a representative
example of an interaction potential.  We also perform an accuracy test
which shows that our method is robust for very large barriers.

The approach presented here could be adapted to shape resonances in
low energy scattering~\cite{robin_lyyra_res}, and to ultra-long-range
Rydberg molecular potentials~\cite{Stanojevic2006,Stanojevic2008}, and
may also prove useful for analyzing threshold
behavior~\cite{robin_nikitin_diss_threshold, robin_heller_suppression}
relevant to ultracold molecules~\cite{Cote1999,byrd2010}, especially
when near-threshold resonances
exist~\cite{ntr-pra-rapid-2014,jost-2015,efimov-2016,Shu_2017}.
Moreover, we are currently investigating the possibility of extending
the phase-amplitude formalism to coupled-channel problems which would
allow studies of Feshbach resonances~\cite{Gacesa2008,Deiglmayr2009}.

\begin{acknowledgments}

  This work was partially supported by the National Science Foundation
  Grant PHY-1806653 and by the MURI U.S. Army Research Office Grant
  No.~W911NF-14-1-0378.
  
\end{acknowledgments}

\appendix

     \section {Proof of the linear independence of the fundamental set
       of solutions $\{\phi^2$, $\chi^2$, $\phi\chi\}$}

    \label{app:proof}

In our previous work~\cite{delta_int} it was shown that if $\phi$ and
$\chi$ are any two solutions of the radial Schr\"odinger
equation~(\ref{eq:radial}), then $\phi^2$, $\chi^2$, and $\phi\chi$
are particular solutions of the envelope equation.  Here we prove that
the triplet $\{\phi^2, \chi^2, \phi\chi\}$ is a basis in the
three-dimensional space of solutions of Eq.~(\ref{eq:rho-linear}), if
$\phi$ and $\chi$ are linearly independent;  specifically, we show
that the linear combination
\begin{equation}\label{eq:rho=abc}
 \rho = a\phi^2+b\chi^2+2c\phi\chi
\end{equation}
vanishes if and only if $a=b=c=0$.  We first use the fact that any
solution of the envelope equation yields an
invariant~\cite{delta_int},
\begin{equation}\label{eq:rho-invar}
Q=\frac 1 2 \rho\rho'' - U\rho^2 -\frac 1 4 (\rho')^2,
\end{equation}
and we employ Eq.~(\ref{eq:rho=abc}) to substitute $\rho$, $\rho'$ and
$\rho''$ in terms of $\phi$ and $\chi$ in the equation above; a
straightforward but tedious manipulation yields
\begin{equation}\label{eq:ab-c2=q2w-2}
Q= (ab-c^2)W^2,
\end{equation}
where $W$ is the Wronskian of $\phi$ and $\chi$.  If $\rho=0$ in
Eq.~(\ref{eq:rho=abc}), we obtain $Q=0$ trivially
from Eq.~(\ref{eq:rho-invar}), while  the linear independence of $\phi$ and
$\chi$ ensures $W\neq 0$, and consequently Eq.~(\ref{eq:ab-c2=q2w-2})
yields
\[
ab=c^2.
\]

We now consider the two possible cases: $c=0$ and $c\neq0$.
In the first case we have $c^2=ab=0$, which implies $a=0$ or $b=0$;
the vanishing of the remaining coefficient ($b$ or $a$, respectively)
follows from our assumption, i.e., $\rho=0$ in Eq.~(\ref{eq:rho=abc}).
For the second case ($c\neq0$, and hence $ab\neq0$), we substitute
$c=\sgn(c)\sqrt{ab}$ in Eq.~(\ref{eq:rho=abc-lin-combi}), and we
obtain
\[
\rho=\pm\Psi^2,
\]
where $\Psi$ is the linear combination
\[
\Psi = \phi\sqrt{|a|}  \pm  \chi\sgn(c)\sqrt{|b|}.
\]
In the two expressions above, the algebraic sign ($\pm$)  is
$\sgn(a)=\sgn(b)$.  Finally,  $\rho=0$ in Eq.~(\ref{eq:rho=abc})
yields $\Psi=0$, and the equation above  implies
$a=b=0$, because $\phi$ and $\chi$ are linearly independent.  This
contradicts the assumption $ab\neq0$ in the second case, which
completes our proof.  Thus, Eq.~(\ref{eq:rho=abc}) with arbitrary
constants $a$, $b$ and $c$ can indeed be
regarded as the general solution of the envelope equation.

  \section{Extending Milne's phase  outside its domain of smoothness}

  \label{sec:extend}

In this appendix we derive a formula for extending the outer phase
$\theta$ into the inner region.  First, the inward propagation of
$\theta$ through the outer region (including the barrier) is
accomplished using the numerical method developed in our previous
work~\cite{delta_int}.  Next, we use Eqs.~(\ref{eq:theta-prime-rho})
and (\ref{eq:rho=abc-lin-combi}) to obtain $\theta(R)$ inside the
inner region ($0<R<R_\text{in}$),
\[
\theta_* - \theta(R) = \int_{R}^{R_\text{in}} \!\! \frac k {\rho(r)} dr
 =  k\!\! \int_R^{R_\text{in}} \frac{dr} {a\phi^2+b\chi^2+2c\phi\chi},
\]
where $\theta_*\equiv\theta(R_\text{in})$ is known.  As explained in
Sec.~\ref{sec:rho}, the integral  cannot be handled numerically
inside the inner region; instead, we tackle it formally.  Making
use of the Wronskian $W=\phi'\chi-\phi\chi'\neq 0$, which is
independent of $r$, we rewrite the integral above,
\begin{equation*}\label{eq:theta=int-W}
\theta_* - \theta(R) =
 \frac k W \! \int_R^{R_\text{in}} \!\!
 \frac{\phi'\chi-\phi\chi'}{a\phi^2+b\chi^2+2c\phi\chi} dr.
\end{equation*}
Next, we define $z(r) \equiv \frac{\phi(r)}{\chi(r)}$ and we change
the integration variable from $r$ to $z$, but we do so only after the
inner region is partitioned in sub-intervals delimited by the nodes of
$\chi(r)$, such that $z(r)$ is a one-to-one mapping inside each
interval.  The change of variable yields
\begin{widetext}
\begin{eqnarray}
\theta_*-\theta(R)
&=&
 k \! \int_R^{R_n} \! \frac{dr}{\rho(r)} +
 k \! \sum_{j=n}^{N_*} \! \int_{R_j}^{R_{j+1}}\frac{dr}{\rho(r)}
\nonumber
\\
 &=& \frac k W \left( \int_{z(R)}^\infty\frac{dz}{az^2 +2cz+b}
              +(N_*-n)\int_{-\infty}^{+\infty}\frac{dz}{az^2 +2cz+b}
              +       \int_{-\infty}^{z_*}\frac{dz}{az^2
                +2cz+b}\right),
\label{eq:theta=int-long}
\end{eqnarray}
\end{widetext}
where $R_1,R_2,\ldots,R_{N_*}$ are the nodes of $\chi$ inside the
inner region, while $R_0=0$ and $R_{N_*+1}=R_\text{in}$ are its
boundaries.  The node $R_n>R$ is the node closest to $R$ inside the
integration domain $[R,R_\text{in}]$.  The upper limit of the last
integral is $z_{*}\equiv\frac{\phi(R_\text{in})}{\chi(R_\text{in})}$.
The new integration variable $z$ in Eq.~(\ref{eq:theta=int-long})
makes it clear that, except for the first and last interval, all other
($N_*-n$) intervals give identical contributions.

Making use of the constraint (\ref{eq:abc-constr-k}), the integral
appearing repeatedly in Eq.~(\ref{eq:theta=int-long}) takes a simple
form,
\[
\frac k W \int\frac{dz}{az^2+2cz+b} = \arctan\left(\frac W k(az+c)\right),
\]
which we now evaluate for each interval.  The contribution of the
first interval is
\[
\frac k W\int_{z(R)}^\infty\frac{dz}{az^2+b+2cz}
 =\frac\pi 2 -\arctan\left(\frac{W}{k}[a\,z(R)+c]\right),
\]
while the last interval yields
\[
\frac k W \int_{-\infty}^{z_{*}}\frac{dz}{az^2+b+2cz} 
   = \arctan\left(\frac W k(az_*+c)\right) +\frac\pi 2.
\]
As mentioned above, the ($N_*-n$) remaining intervals give identical
contributions; namely, for $n\leq j\leq N_*-1$, we have
\[
k\int_{R_j}^{R_{j+1}}\frac{dr}{\rho(r)}
 = \frac k W \int_{-\infty}^{+\infty}\frac{dz}{az^2+b+2cz} =\pi.
 \]

Finally, we add the contributions from all intervals to obtain the
outer phase $\theta$ inside the inner region,
\begin{eqnarray}
  \theta(R)
  &=&
  \arctan\left(\frac{W}k[a\,z(R)+c]\right),
           \label{eq:theta-at-R}
  \\
           &-& \arctan\left(\frac W k(c+az_*)\right)+\theta_* - \pi(N_*-n+1).
      \nonumber
\end{eqnarray}
This result is of key importance, as it yields the scattering phase
shift; see Sec.~\ref{sec:delta}.

\section{
  Choosing the location of the matching point}

\label{app:thor}

For energies above the barrier, $R_\text{top}$ is a convenient
location for the matching point, while for scattering energies below
the top of the barrier the matching conditions are imposed at
$R_\text{in}$.  However, for $E<E_\text{top}$, the matching point can
be placed anywhere within the classically forbidden region under the
barrier, despite the fact that in Sec.~\ref{sec:match} we argued that
the matching point be located at the turning point $R_\text{in}$ (see
Fig.~\ref{fig:Veff}).  $R_\text{in}$ is a necessary choice for the
matching point only if the phase-amplitude approach is restricted to
the outer region; see Sec.~\ref{sec:match}\@.  Indeed, if the
phase-amplitude method is also used in the inner region, the matching
point need no longer be kept at (or near) $R_\text{in}$.  The freedom
to relax the location of the matching point stems from the fact that
the inner solutions $\phi$ and $\chi$ can be parametrized in terms of
the inner envelope $\varrho$ and phase $\beta$, as shown in
Sec.~\ref{sec:opt}\@.  Accordingly, the solutions~(\ref{eq:abc-raw})
of the the matching equations~(\ref{eq:match}) are expressed entirely
in terms of phase-amplitude quantities and remain highly accurate if
the matching point (which we now denote $R_*$) is moved between
$R_\text{in}$ and $R_\text{out}$ (the outermost turning point).

Although the scaled coefficients introduced in Eq.~(\ref{eq:abc-tilde})
are formally independent of the matching point, their simplified
expressions~(\ref{eq:abc-clean}) are no longer  independent
of $R_*$.  To clarify this aspect, we now analyze the $R_*$ dependence
of the inner-region phase $\beta_\text{full}$ in
Eq.~(\ref{eq:beta-full}) to show that for energies sufficiently lower
than $E_\text{top}$ the phase $\beta_\text{full}(R_*)$ is practically
independent of the matching point.  Specifically, we make use of the
definition (\ref{eq:cot-eta}) to evaluate the derivative
$\eta'=d\eta/dR_*$, while from Eq.~(\ref{eq:beta}) we have
$\beta'=q/\varrho$.  Taking advantage of the
invariant~(\ref{eq:rho-invar}) with $Q=q^2$ and $Q=k^2$ for $\varrho$
and $\rho$, respectively, we obtain
\[
\beta'_\text{full} =\beta'+\eta' = \frac{k^2\varrho}{\rho^2 q} \sin^2\eta,
\]
which is vanishingly small for $E$ sufficiently lower than
$E_\text{top}$.  Indeed, if $R_*=R_\text{in}$, we have $\rho(R_*)\ggg
1$, which ensures $\beta'+\eta'\approx0$.  If $R_*$ is shifted away
from $R_\text{in}$, then $\varrho$ increases while $\eta$ and $\rho$
decrease; from Eq.~(\ref{eq:cot-eta}) we have $\eta\sim\frac
1{\varrho}$ when $\varrho\ggg 1$, and we  find
\[
\beta'_\text{full} =\beta'+\eta' \sim \frac{k}{\rho^2\varrho}\approx 0.
\]
Therefore, we have
\[
\beta_\text{full}(R_*) = \beta(R_*)+\eta(R_*) \approx \text{constant,}
\]
which justifies our interpretation of $\beta+\eta=\beta_\text{full}$ as
the full phase accumulated at short range, including the contribution
from the barrier region; indeed, when $R_*$ is near $R_\text{out}$, we
have $\eta\approx 0$, and thus
\[
\beta_\text{full}(R_*) \approx \beta(R_\text{out}),
\quad\text{for } R_\text{in}\leq R_*\leq R_\text{out}.
\]

Finally, we remark that $R_*=R_\text{top}$ is a convenient choice for
the matching point for all energies (below and above the barrier).  In
general, the matching point can be energy dependent, e.g., the turning
point $R_\text{in}(E)$.  Therefore, in order to ensure the quantities
$\beta$, $\eta$, $u$ and $\varepsilon$ introduced in
Sec.~\ref{sec:opt} have a well defined energy dependence, the matching
point $R_*(E)$ must be chosen such that it is a well behaved function
of energy.

\bibliography{barrier} 

\end{document}